\def\MYJOURNAL{1} 

\if 0\MYJOURNAL%
\documentclass[3p, preprint, sort&compress]{elsarticle}
\usepackage{setspace}
\else \if 1\MYJOURNAL%
\documentclass[9pt,journal,compsoc]{IEEEtran}
\usepackage{geometry}%
\usepackage[numbers,sort&compress]{natbib}
\usepackage{setspace}

\fi
\fi

\usepackage{color, xcolor, colortbl}
\usepackage{hyperref}
\usepackage{amsfonts}
\usepackage{amsmath}
\usepackage{amssymb}
\usepackage{graphicx}
\usepackage{url}
\usepackage{fp}
\providecommand{\U}[1]{\protect\rule{.1in}{.1in}}

\usepackage{enumitem}
\usepackage[referable,para]{threeparttablex}
\usepackage{footnote}
\usepackage{booktabs}  

\usepackage{mwe}
\usepackage[percent]{overpic}

\graphicspath{{./Graphics/}}

\usepackage{listings}
\definecolor{dkgreen}{rgb}{0,.6,0}
\definecolor{dkblue}{rgb}{0,0,.6}
\definecolor{dkyellow}{cmyk}{0,0,.8,.3}

\lstdefinestyle{customphp}{
  language        = php,
  basicstyle      = \small\ttfamily,
  keywordstyle    = \color{dkblue},
  stringstyle     = \color{red},
  identifierstyle = \color{dkgreen},
  commentstyle    = \color{gray},
  emph            =[1]{php},
  emphstyle       =[1]\color{black},
  emph            =[2]{if,and,or,else},
  emphstyle       =[2]\color{dkyellow}}

\lstdefinestyle{customc}{
  breaklines=true, breakindent=20pt,
  frame=leftline,
  numbers=left,
  language=C, numberstyle=\tiny, numbersep=10pt,
  showstringspaces=false,
  basicstyle=\footnotesize\ttfamily,
  keywordstyle=\bfseries\color{green!40!black},
  commentstyle=\itshape\color{purple!40!black},
  identifierstyle=\color{blue},
  stringstyle=\color{orange},
  captionpos=t
}

\hypersetup{
  colorlinks,
  citecolor=blue,
  filecolor=blue,
  linkcolor=blue,
  urlcolor=blue
}

\usepackage{fourier}
\DeclareMathAlphabet{\mathcal}{OT1}{pzc}{m}{it}
\DeclareSymbolFont{letters}{OML}{cmm}{m}{it}

\usepackage{tikz}
\usepackage{float}
\usepackage{pgf}
\usepackage{pgfplots}

\usetikzlibrary{calc}
\usepgflibrary{shapes}
\usetikzlibrary{er}
\usetikzlibrary{arrows,snakes,backgrounds}
\usetikzlibrary{decorations.text}

\usepgfplotslibrary{external}
\usetikzlibrary{spy}

\def\getangle(#1) (#2)#3{%
  \begingroup%
  \pgftransformreset%
  \pgfmathanglebetweenpoints{\pgfpointanchor{#1}{center}}{\pgfpointanchor{#2}{center}}%
  \expandafter\xdef\csname angle#3\endcsname{\pgfmathresult}%
  \endgroup%
}

\pgfplotsset{compat=1.11}
\usetikzlibrary{calc,trees,positioning,arrows,chains,shapes.geometric,%
  decorations.pathreplacing,decorations.pathmorphing,shapes,%
  matrix,shapes.symbols}

\tikzset{
  >=stealth',
  punktchain/.style={
    font=\scriptsize,
    rectangle,
    rounded corners,
    draw=black, thick,
    text width=10em,
    minimum height=1em,
    text centered},
  line/.style={draw, thick, <-},
  element/.style={
    tape,
    top color=white,
    bottom color=blue!50!black!60!,
    minimum width=8em,
    draw=blue!40!black!90, very thick,
    text width=10em,
    minimum height=1em,
    text centered},
  every join/.style={->, thick,shorten >=1pt},
  decoration={brace},
  tuborg/.style={decorate},
  tubnode/.style={midway, right=2pt},
}

\if 0\MYJOURNAL%
\tikzset{
  PIXEL/.style={
    font=\fontsize{4}{3.6}\selectfont,
    text width=9em,
    minimum height=1em,
    text centered
  }
}
\else \if 1\MYJOURNAL%
\tikzset{
  PIXEL/.style={
    font=\tiny,
    text width=8em,
    minimum height=3em,
    text centered
  }
}
\fi
\fi

\usepackage{multirow}
\usepackage{subfigure}
\newcommand{\libname}{JSOL}
\definecolor{mygreen}{rgb}{0,0.6,0}
\definecolor{mygray}{rgb}{0.5,0.5,0.5}
\definecolor{mymauve}{rgb}{0.58,0,0.82}

\definecolor{darkgray}{rgb}{.4,.4,.4}
\definecolor{purple}{rgb}{0.65, 0.12, 0.82}

\lstdefinelanguage{JavaScript}{
  keywordstyle=\color{blue}\bfseries,
  ndkeywords={class, export, boolean, throw, implements, import, this},
  ndkeywordstyle=\color{darkgray}\bfseries,
  identifierstyle=\color{black},
  sensitive=false,
  comment=[l]{//},
  morecomment=[s]{/*}{*/},
  commentstyle=\color{purple}\ttfamily,
  stringstyle=\color{black}\ttfamily,
  morestring=[b]',
  morestring=[b]"
}

\lstdefinestyle{customJS}{
  language=JavaScript,
  extendedchars=true,
  basicstyle=\footnotesize\ttfamily,
  showstringspaces=false,
  showspaces=false,
  numbers=left,
  numberstyle=\tiny,
  numbersep=0pt,
  tabsize=2,
  breaklines=true,
  showtabs=false,
  captionpos=b
}

\usepackage{listings, xcolor}
\usepackage{tabularx}

\begin{document}

\if 0\MYJOURNAL%
    \input{SecAbstract-PR.tex}
\else \if 1\MYJOURNAL%
    \title{JSOL:\\JavaScript Open-source Library for Grammar of Graphics}

\author{%
  Yousef, Waleed A.\textsuperscript{a},~\IEEEmembership{Senior Member,~IEEE,}~\thanks{Yousef, Waleed
    A., is an associate professor, computer science department, faculty of computers and
    information, Helwan University, Egypt, \url{wyousef@fci.helwan.edu.eg}}
  Mohammed, Hisham E.\textsuperscript{a},~\thanks{Mohammed, Hisham E., B.Sc., \url{hishamelamir001@gmail.com}}
  Naguib, Andrew A.\textsuperscript{a},~\thanks{Naguib, Andrew A., B.Sc., \url{andrew@fci.helwan.edu.eg}}

  Eid, Rafat S. \textsuperscript{a,b},~\thanks{Eid, Rafat S., B.Sc., \url{Raafat.Sobhy5@gmail.com}}
  Emabrak, Sherif E.\textsuperscript{a,b},~\thanks{Embarak, Sherif E., B.Sc., \url{sherif.embarak@gmail.com}}
  Hamed, Ahmed F.\textsuperscript{a,b},~\thanks{Hamed, Ahmed F., B.Sc., \url{afouad1511@gmail.com}}
  Khalifa, Yusuf M.\textsuperscript{a,b},~\thanks{Khalifa, Yusuf M., B.Sc., \url{Ymk.hassan@gmail.com}}
  AbdElrheem, Shrouk T.\textsuperscript{a,b},~\thanks{AbdElrheem, Shrouk T., B.Sc., \url{doha55tarek@gmail.com}}
  Awad, Eman A.\textsuperscript{a,b},~\thanks{Awad, Eman A., B.Sc., \url{emanawad1995@gmail.com}}
  Gaafar, Sara G.\textsuperscript{a,b},~\thanks{Gaafar, Sara G., B.Sc., \url{saragaafar1000@gmail.com}}
  Mamdoh, Alaa M.\textsuperscript{a,b},~\thanks{Mamdoh, Alaa M., B.Sc., \url{alaamamdouh800@gmail.com}}
  Shawky, Nada A.\textsuperscript{a,b},~\thanks{Shawky, Nada A., B.Sc., \url{nadashawky2000@gmail.com}}

  \begin{center}
    \hfil\includegraphics[height=0.17\textheight]{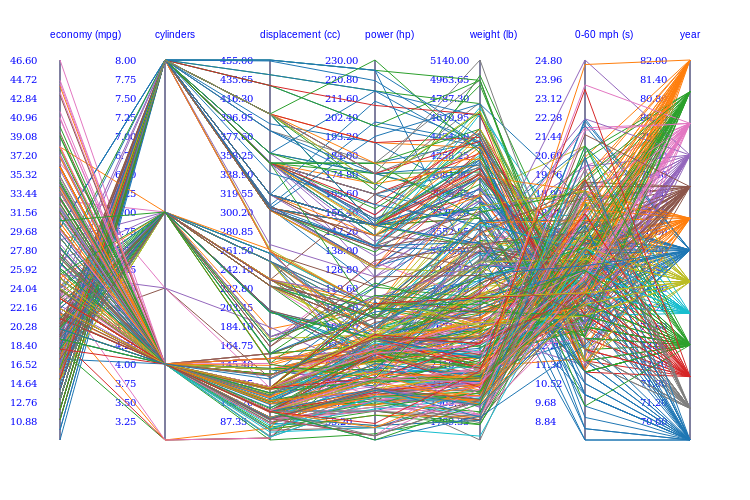}\hfil\includegraphics[height=0.17\textheight]{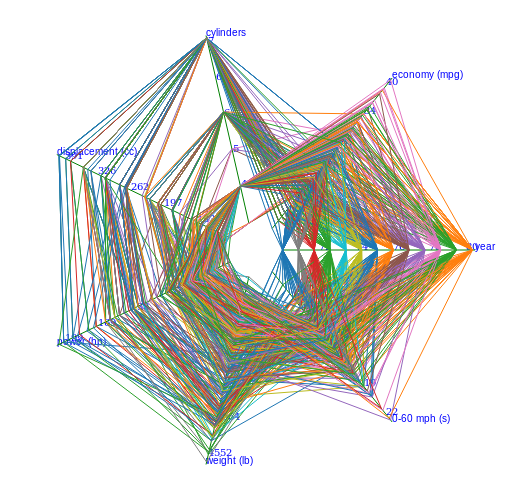}\hfil\includegraphics[height=0.17\textheight]{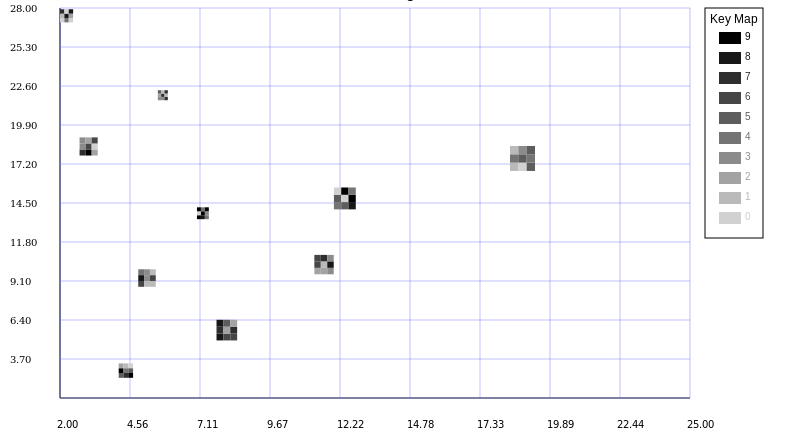}\hfil
    {\footnotesize Three different figures produced by the JSOL library: parallel coordinate (left),
      circular parallel coordinates (center), and very customized multi dimensional plot, where
      $k^2$ dimensions are the gray level of a created $k \times k$ image and the other 2 dimensions
      are the image $x$-$y$ location on a 2-D scatter plot}
  \end{center}

  \thanks{\textsuperscript{a}Human Computer Interaction Laboratory (HCI Lab.), Egypt.}
  \thanks{\textsuperscript{b}These authors contributed equally to the manuscript as the fourth author and names are
    ordered alphabetically according to the family name.}  }

  \maketitle

  \begin{abstract}
    In this paper, we introduce the JavaScript Open-source Library (\libname), a high-level grammar for
representing data in visualization graphs and plots. \libname~perspective on the grammar of graphics
is unique; it provides state-of-art rules for encoding visual primitives that can be used to
generate a known scene or to invent a new one. \libname~has ton rules developed specifically for
data-munging, mapping, and visualization through many layers, such as algebra, scales, and
geometries. Additionally, it has a compiler that incorporates and combines all rules specified by a
user and put them in a flow to validate it as a visualization grammar and check its
requisites. Users can customize scenes through a pipeline that either puts customized rules or comes
with new ones. We evaluated \libname~on a multitude of plots to check rules specification of
customizing a specific plot. Although the project is still under development and many enhancements
are under construction, this paper describes the first developed version of \libname, circa 2016,
where an open-source version of it is available~\cite{JSOL2016}. One immediate practical deployment
for JSOl is to be integrated with the open-source version of the Data Visualization Platform (DVP)
\citep{Yousef2019DVP-arxiv}


  \end{abstract}

  \newcommand{\sep}{,~} 


    \fi
\fi

\section{Introduction}\label{sec:introduction}
The most popular visualization steps, generally speaking, and particularly for data visualization is as easy as to get your dataset into a visualization library (or system), defining your charting information, and then you have the chart you need. These steps could be implemented in a multitude of ways; while the technology is rising, most of the software is attempting to cope up and contain the top used graphs and visualization scenes and also modifying the way to generate these graphs by using a user interface to make it milder to construct graphs with few clicks. This methodology has the advantage when it comes to quick and easy use, but the disadvantage is that this software is restrictive and does not allow internal customizations and modification. The other alternative is to use a charting system such as vector drawing which permits the user to customize every single piece of the graph, the interest here is that the user can build the chart freely without any restrictions. However, it is troublesome to use for most of the users and consumes a considerable amount of time to build a full scene; thus, the need arises for a visualization library that combines all of the merits.

\subsection{Grammar of Graphics and Other Systems} 
Grammar of Graphics (GoG) is a set of rules put together to create a scene that expresses the data; divided into two categories, \textit{Low-Level} grammar, that is used to customize each piece of the visualization scenes; mainly used for exploratory data analysis as their primitives offer fine-grained control ---like scenes used in analysis tools. By way of illustration, \cite{2011-d3} and \cite{Bostock2009Protovis}. On the other hand, \textit{High-Level} grammar is used to make traditional visual plots expeditiously. It is useful to users and analysts for rapid development. Examples include \cite{ggplot2}, \cite{vega-lite}, and grammar-based systems such as \cite{tableau}. High-level grammars are more prevalent as users prefer conciseness over expressiveness. Furthermore, In contrast to the \textit{low-level} grammars, the \textit{high-level} grammars use default values to resolve ambiguities of visualization, hence development is comfortable for analysts and developers, At last there is some libraries that are in the middle level such \textbf{protovis}, those libraries tries to combines the ease of use and the non-restrictive methodology to generate a middle-ware level that combine the advantages of each of low-level and high-level.\\
Choosing between the low-level or high-level is not easy, user should take in consideration some concepts, such as the time-consumption or in other words how long it takes to build the scene? and thats called "\textit{efficiency}", another concept is do the user have the knowledge to build the scene or is the knowledge exists to make the user learn how to use the system/library and that's called "\textit{accessibility}", at last can the user build the scene or not? and this is the "\textit{expressiveness}", every user should take these concepts in consideration when it comes to generate a beautiful plots, sometimes it confuses the users when it comes to choose between those levels and that's the main reason for inventing the middle-level and the hybrid-libraries.

\subsection{Why Another Library?}\label{sec:why-another-library}
\libname~belongs to high-level grammars because of the compiler it presents, which has built-in default values passed to low-level layers when not specified by the user. As a result, rapid development accomplished. Moreover, \libname's layers are a real representation of low-level grammars: representing customization for each visualization component.

\bigskip

\libname's layers are built using \textit{JavaScript} scripting language with its compiler which uses \textit{JSON} (JavaScript Object Notation) to interpret the user's definition of specifications. Consequently, it entirely works and runs in web browsers that support JavaScript. Specifically, \libname\space runs in \textit{HTML5 canvas} elements that allow for high
performance.

Users may think that the debugging will be problematic; however, that is not the case because of \libname's compiler having an inherent error handling layer that provides the user with a proper debugger and informative error/warning messages.

There are some objectives we cannot miss in any visualization library,
which are:
\begin{itemize}
  \item\textbf{Performance}: high-level abstraction may limit the user's ability to generate fast
  visualization scenes. However, building charts on top of HTML5 Canvas elements allow
  \libname\space to use \textit{GPU} acceleration to speed up the process, even though we will be
  shifting the responsibility of data representation and transformation to the user and no longer
  treat it as our concern.
  \item\textbf{Debugging}: trial and error is a fundamental part of the development and learning
  processes; accessible tools must be designed to support debugging when errors occur. As
  \libname\space was built using JavaScript, it allows users to use various types of debugging
  tools, on account of the built-in test/debug layer that enhances the package's efficiency.
\end{itemize}

\bigskip

\libname~comprises low-level layers, called \textit{modules} or \textit{kernels}, that are made specially to carry all the burden of a given task. Visualization primitives alternatively named \textit{marks} in \cite{vega-lite} or shape in \cite{2011-d3}, provide geometries or shapes for chartings, such as a bar, point, and an arc. Another layer is the data encoding layer, also called \textit{scale}, it allows mapping data points to encoded pixels to bring data to life, for using a particular scale layer you need to generate a reference via the \textit{Axes} layer which allows creating \textit{Cartesian} or \textit{Polar} coordinated charts and plots. There is also a layer for data munging, called \textit{algebra}, made for practicing operations on data such as \textit{join} (left and right), \textit{cross}, and \textit{nesting} which allows visualizing in higher dimensions as we will later see in the examples section. Over and above, there are statistics for making summary statistical operations on data, such as \textit{mean}, \textit{std}, and so on. An expansive and comprehensive description of the library illustrated in Sec \ref{sec:brief-insight-libn}.

\bigskip

The \libname~compiler synthesizes and combines all low-level rules and specification gained from the other layers within respect for a given data and validates the users' rules through all layers in \libname\space using the handler which manages default values for visualization primitives. In a wide range of examples, we will confirm how the compiler takes a tremendous advantage of the lower-level encoding and visualization layer of the library to bring a high-level specification to visualizations, and later will demonstrate how the compiler works and what are the minimum values must be given to make visual plots.

\bigskip

On the one hand, visualization systems considered a subordinate of graphical ones. Moreover, they are the responsible entities to produce and process representation of data graphically and its interaction to gain insight into the data. On the other hand, graphical systems used to generate drawings in general, offering the utmost flexibility. They also have different types (discussed in Sec. \ref{sec:related-work}). Nevertheless, primarily, they were not tailored for visualization purposes.


\section{Background and Related Work}\label{sec:related-work}
The original author of grammar-based visualization concept is \cite{Wilkinson}, changing the way
scientists and developers think, as well as inspiring them. Stanford's team in \cite{Polaris} has
planted the seed of the grammar of graphics' software, after which gobs of systems developed ---such
as Tableau in \cite{tableau}, ggplot2 in \cite{ggplot2}; although their user preferences'
customization is limited, they brought in abstraction of data models, graphical geometries, visual
encoding channels, scales, and guides (e.g., axes and legends), yielding more expressive design
space. \libname's architecture heavily influenced by these works; inheriting from Wilkinson's
grammar and components; rendering basic graphical scenes using these components and the grammar that
makes these components generate a full scene. The component instance could be a visual channel such
as \textit{position}, \textit{color}, \textit{shape}, and \textit{size}, may also include common
data transformations, as a sample, binning, aggregation, sorting, and filtering. The grammar is the
validation step that is responsible for making these components work in flow and mapping data
attributes to its component in the scene.

\subsection{Specification}
Visualization libraries are of a distinct number of architectures' types: 
\begin{itemize}
  \item \textbf{Hierarchical:} Layers and components are implemented in a hierarchical view, so that
  when a new scene graph introduced, authors build it in layers employing the components. In case
  that the graphic depends on the same component they will be linked.
  \item \textbf{Parallel:} Building layers and components in two parallel independent levels, so
  when a scene graph proposed users use the component given by the system to implement it or to
  customize their graphics.
  \item \textbf{Hybrid:} It is a combination of the previous
  couple of types. The compiler and layers are hierarchical; however, layers' design is parallel
  (\libname\space adopts this type).
\end{itemize}

The compiler added on the top layer makes \libname\space a declarative \textit{domain specific
  language (DSL)} for visualization design; by decoupling specification from execution details,
declarative systems allow users to focus on specifying their application domain without limiting
their abilities to customize.

\cite{vega-lite} and \cite{Bostock2009Protovis} followed the same approach of the DSL compilation criterion;
however, they use a declarative framework for mapping data to visual elements. Nevertheless,
\libname\space does not strictly impose a toolkit-specific lexicon of graphical marks; instead,
\libname\space directly maps data attributes to the HTML5 canvas element.

\subsection{Other Libraries}\label{sec:comparative-study}
\subsubsection{GGPlot2}\label{sec:ggplot2}
In \cite{ggplot2}, there is excessive concentration on low-level details which makes plotting a
hassle (e.g., drawing guides). However, it provides a dominant model for graphics that makes it easy
to produce complex multi-layered scenes. You can hardly find demerits in \textit{ggplot2}, but one
of them is the language it uses, \textit{R}, as discussed earlier, is a programming language that
uses a specific interpreter, making it laborious for users to install for developing some simple
visualization scenes, yet it is normal for R audience to use. Furthermore, R does not support
variables' or methods' labeling which makes users struggle when using it. Besides, it only works on
CPUs, making it slow against other libraries.

\subsubsection{D3}\label{sec:d3}
Many scientists, researchers, developers, and even programs make use of \textit{D3}, it is nearly
suitable for each user, but there are few crucial drawbacks; it is entirely low-level visualization
grammar which is hard for novice users to manage, one extra drawback appears when working on an SVG
element in big data, which is generating a ton of elements that may break browsers as a result of
the load added on the browser's bag.

\subsubsection{Protovis}\label{sec:protovis}
We conducted a comparison between \cite{Bostock2009Protovis} and \libname\space in Table \ref{table:1},  \cite{Bostock2009Protovis} composes custom views of data with simple marks such as pies (\includegraphics*[height=\fontcharht\font`\B]{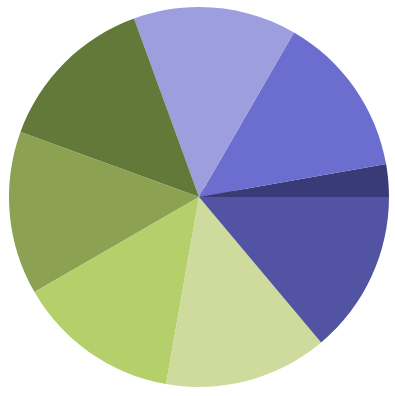}) and dots. Unlike low-level graphics libraries that quickly become tedious for visualization, Protovis defines marks through dynamic properties that encode data, allowing inheritance, scales and layouts to simplify construction. However it is no longer under active development and that make it unbearable for different types of users.

\subsubsection{Vega}\label{sec:vega}
In a previous section we saw that Vega specification is simply a JSON object that describes an interactive visualization, and that may appear akin to \libname. However vega uses \cite{2011-d3} as a backend engine to produce and provide a SVG visulization components given a user-grammars. On the other hand, \libname\space specification may be cross-compiled to provide a reusable visualization component, of course given the user input grammars.

\subsubsection{Vega-Lite}\label{sec:vega-lite}
We conducted a comparative study between \cite{vega-lite} and \libname\space in Table \ref{table:1},
these two libraries are chosen for a particular purpose which is that the users may confound that
they are equal, true that they are similar in attributes, but not in drawbacks. \libname\space
specifically developed for making a complete detailed scene, that is why it collects all
specification that developers or scientists seek to configure their scene.
\begin{table}[tbh]
  \centering
  \caption{Comparative Study}\label{table:1}
  \renewcommand{\arraystretch}{1.0}
  \begin{tabular}{|l|l|l|l|l|l|}
    \hline
    \multicolumn{3}{|l|}{Grammar of Graphics layers} & Protovis & Vega & \libname  \\ \hline
    \multicolumn{3}{|l|}{Transformation layer} & Done & Done & Done \\ \hline
    \multicolumn{3}{|l|}{Data layer} & Done & Done & Done \\ \hline
    \multirow{17}{*}{\begin{tabular}[c]{@{}l@{}}Geometry \\ layer\end{tabular}} & \multicolumn{2}{l|}{Point} & Done & Done & Done \\ \cline{2-6} 
    & \multirow{4}{*}{Bar} & bar chart & Done & Done & Done \\ \cline{3-6} &  & \begin{tabular}[c]{@{}l@{}}Stacked bar\\ chart\end{tabular} & Done & Done & Done \\ \cline{3-6}
    &  & Histogram & Done & Done & Done \\ \cline{3-6} &  & 
    \begin{tabular}[c]{@{}l@{}}Vertical bar\\ chart\end{tabular}  & Done & Done & Done \\ \cline{2-6}
	& \multirow{2}{*}{Area} & Area chart & Done & Done & Done \\ \cline{3-6}
    &  & \begin{tabular}[c]{@{}l@{}}Stacked Area \\ chart\end{tabular} & Done & Done & - \\ \cline{2-6}
	& \multicolumn{2}{l|}{Text} & Done & Done & Done \\ \cline{2-6}
	& \multicolumn{2}{l|}{Line} & Done & Done & Done \\ \cline{2-6}
	& \multicolumn{2}{l|}{Marks} & Done & Done & Done \\ \cline{2-6}
	& \multicolumn{2}{l|}{HLine} & Done & Done & Done \\ \cline{2-6}
	& \multicolumn{2}{l|}{VLine} & Done & Done & Done \\ \cline{2-6}
	& \multicolumn{2}{l|}{Pie chart} & Done & Done & Done \\ \cline{2-6}
	& \multicolumn{2}{l|}{Arc chart} & Done & Done & Done \\ \cline{2-6}
	& \multicolumn{2}{l|}{Picture} & Done & Done & Done \\ \cline{2-6}
	& \multirow{2}{*}{Tick} & Dot plot & Done & -& - \\ \cline{3-6}
	& & Strip plot & Done & - & - \\ \hline
    \multicolumn{3}{|l|}{Scale layer} & Done & Done & Done \\ \hline
    \multirow{6}{*}{Axes Layer} & \multicolumn{2}{l|}{Cartesian coordinate} & Done & Done & Done \\ \cline{2-6}
    & \multicolumn{2}{l|}{Coordinate equal} & Done & - & Done \\ \cline{2-6}
    & \multicolumn{2}{l|}{Coordinate flip} & Done & - & Done \\ \cline{2-6}
    & \multicolumn{2}{l|}{Coordinate polar} & Done & - & Done \\ \cline{2-6}
    & \multicolumn{2}{l|}{Parallel coordinate} & Done & Done & Done \\ \cline{2-6}
    & \multicolumn{2}{l|}{Polar parallel coordinate} & - & - & Done \\ \hline
    \multicolumn{3}{|l|}{Aesthetics layer} & Done & Done & Done \\ \hline
  \end{tabular}
\end{table}


\lstset{
	string=[s]{"}{"},
	stringstyle=\color{blue},
	comment=[s]{:\ "}{"},
	commentstyle=\color{red},
	basicstyle=\footnotesize,
	frame = single,
	framexleftmargin=-2pt,
	framexrightmargin=-10pt,
}

\section{The \libname}\label{sec:brief-insight-libn}
\libname\space amalgamates graphics' grammar with a state-of-art compilation process. Throughout
this section, we will cover how a simple scene is generated and constructed, together with how data
processed; besides, the layers' design.

\subsection{Unit Specification}
Doubtlessly, a scene must have a data variable. After all, that is what going to be a graphic. There
are also some customization parameters ---such as transformation, geometries, properties, and a set
of encodings. Transformation layer is responsible for applying filters and aggregation. After which,
the geometry layer visually encodes the incoming input.

\lstinline[style=customJS]{scene := (data, tansformations, geometries, properties)}

Defining \textit{properties} is optional; however, it is significant when it comes to details; as a
way of illustration, imagine the case where a user wants to declare points' color in the scene or
the type of the data variable (e.g., CSV). \textit{Scale} is also necessary as it determines how
data attributes mapped to traits of geometries. \textit{Axes} enhances the readability of scales.

\lstinline[style=customJS]{properties := (geometries, data, functions, scale, axes, guide)}

\subsection{Layers of~\libname}
As the unit specification complexity issue is evolving daily, layers built with simplicity in mind. Several parts of the kernel is the same as \cite{2011-d3}'s, the other parts modified to flatter the interaction between the user and the library. We will discuss each layer's structure and role.

\subsubsection{Data Layer}
As the layer's name states, its role is to read data from various sources, for example, flat files (e.g., CSV and TSV), and open-standard files (e.g., JSON). Data could be an array of arbitrary values, numbers, strings,
or objects. After reading the data source, the layer stores it in a pre-defined data structure for easily referencing through other layers; the data structure also supports CRUD operations. As shown in the example below, each dataset has a \textit{name} for the consistency of multiple dataset loading, the \textit{values} parameter is the source that contains the dataset, the \textit{format} defines the source type.

\begin{lstlisting}
	"data": [{
		"name": "troops",
		"values": "troops.csv",
		"format": { "type": "csv" }
		}, {
		"name": "cities",
		"values": "cities.csv",
		"format": { "type": "csv" }
	}]
\end{lstlisting}

\subsubsection{Transformation Layer}
This layer's responsibility is to perform analytical operations on data
gathered by the data layer, and it helps the user to perform many
transformations including filtering and grouping. Executing these
operations is necessary to optimize the processing time. The layer
expects that its input comes from the data layer to do a valid procedure on the dataset. It consists of two sub-layers taken from \cite{Wilkinson} which are \textit{variables} and \textit{algebra}.

\begin{lstlisting}
"transform":[{
  "lang": "R",
  "function": "fibonnaci",
  "properties": {
    "data" : "static",
    "length" : 20,
    "field" : "x",
    "name": "fibonnaci_x"
  }
}]
\end{lstlisting}

\begin{figure}[tbh]
	\centering
	\includegraphics[width=0.5\textwidth]{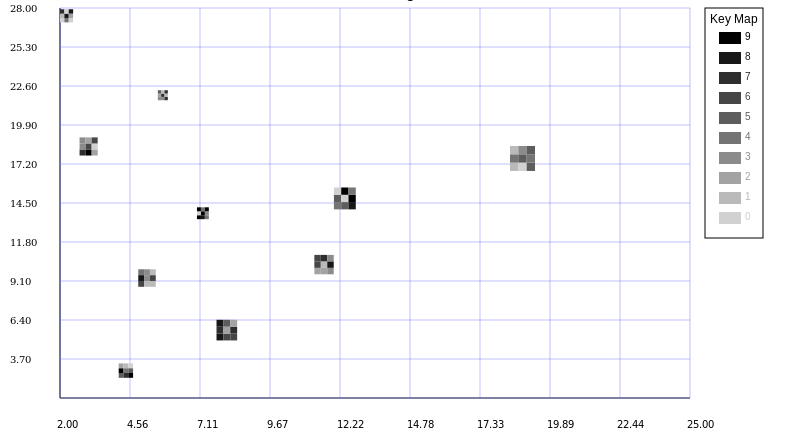}
	\caption{nesting caption}
	\label{fig:nesting}
\end{figure}

\paragraph{High-dimensional Spaces}
Living in a 3-D world restricts us from visualizing structures in high-dimensional spaces. The curse of dimensionality, as called by \cite{bellman}, has been an impediment for so long with various solutions; however, we will only recall \textit{nesting} as embracing all solutions is currently out of our scope.  Nesting is a way to circumvent the challenge, in which we represent any two dimensions of the data on the X-axis and Y-axis. Nevertheless, points designed by these axes are separate charts (e.g., pie, bar, image and so on); in \libname, we use a 9-block greyscale image in which each block's greyscale level is equal to the dimension's value \textit{(see fig. \ref{fig:nesting})}.

\subsubsection{Scales Layer}
A visual encoding is called \textit{scale}. As to draw data in a scene, we need to map data values
to their corresponding geometries. This layer is taken from \cite{2011-d3}, supports both ordinal
and quantitative (linear, logarithmic, exponential, quantile) values.

\begin{lstlisting}
"scales": [{
  "name": "yscale",
  "type": "linear",
  "range": {
    "type": "range",
    "value": "height"
  },
  "domain": {
    "data": "crimea",
    "field": "economy (mpg)"
  }
}]
\end{lstlisting}

\subsubsection{Statistics Layer}
The use of the statistics layer is optional since applying statistical functions is not the central
objective of each user. It is managed using the R programming language which gives us a handicap
while comparing to other libraries as R is a user-friendly language built specially for statistical
modeling and inference, making it light to execute any statistical function on data in a few lines
of codes. Moreover, it has a knowledgeable community and rich documentation.

\subsubsection{Geometry Layer}
\libname~implements the same \texttt{d3.svg.shape} element provided by \cite{2011-d3} in an HTML5 canvas element that is suitable for charting, providing the power of computational speed supported from \textbf{WebGL}; the arc, for example, builds elliptical arcs such as pie (\includegraphics*[height=\fontcharht\font`\B]{basic_pie}), donut and cox-comb (\includegraphics*[height=\fontcharht\font`\B]{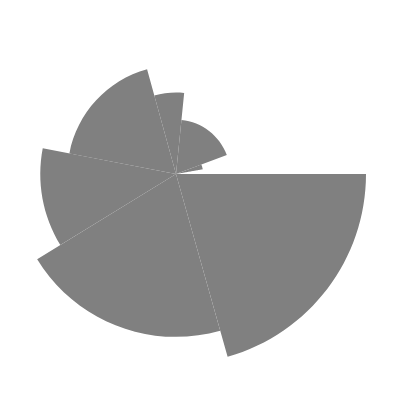}) charts via formulating arbitrary data to paths. Typically, this function bounded to the arc attribute, note that the radius and angles of the arc can be specified both as constants or callback functions. Additional shapes provided for areas, lines, mark symbols, etc.

\begin{lstlisting}
"geom" : [{
  "type" : "Point",
  "data" : "static",
  "properties" : {
    "x" : "xscale",
    "y" : "yscale",
    "fillColor" : "zscale"
  }
}]
\end{lstlisting}

\subsubsection{Axes Layer}
The layer is a crucial step in the graphical scene since it maps the scale to a meaningful form that is comprehensible by human eyes. Axes visualize spatial scale mappings using \textit{ticks}, \textit{grid lines}, and \textit{labels}. \libname\space supports lot of axis based on a given scale, and currently supports axes for Cartesian (rectangular) and Polar coordinates.

\begin{lstlisting}
"axes": [{
  "type": "x",
  "data": "static"
  "field": "x"
  "orient" : "bottom"
  "grid" : true
}, 
{
  "type": "y",
  "scale": "yscale"
}]
\end{lstlisting}

\subsubsection{Guide Layer}
Both \textit{guides} and \textit{axes} visualize scales; but guides aid interpretation of scales
with ranges such as colors, shapes, and sizes, whereas axes aid interpretation of scales with
spatial ranges. Similar to scales and axes, guides can be defined either as a top-level or low-level
visualization.

\begin{lstlisting}
"guides": [{
  "type": "legend",
  "domain": { ... },
  "properties": {
    "title": { 
      "name": "key map"
    },
    "position": { ... }
  }
}]
\end{lstlisting}

\subsection{Compiler of \libname} 
The \libname's compiler is made purposefully to transform the library from low-level visualization library to a high-level one, making it effortless for users to interpret their parameters to build imaginative scenes, the compiler also has predefined value for each parameter; consequently, the headache of passing values uprooted.

The user’s specifications pass through several phases or stages
to become a beautiful chart, the first one is \textit{scanning} user’s
specifications and divide them into sections basing on their’s role; \textit{parse}, which is subject to prepare the low-level representation and fill the missing parameters; \textit{linking}, where layers' objects attached to each other, and finally \textit{assemble} is where the full chart comes alive.\\

In \libname~compilation process the first stage is scanning, here where the users' specification been parsed and validated, \libname~ provides tons of rules, for example each type of scales like linear scale is considered a rule, can be applied by users and developers each of these rules follow some validation steps so we should certain that the \libname~ users will follow it ensure that the scene will be generated. For example some geometry components must have a color pallets, so if the user forgot to set it the compiler must set it to a default color pallet and so on.

Second things second is connecting phase, after the user's specifications are validated, this phase is responsible for generating these specifications, by generating we means that should transform the user's specifications to layers executable classes, function and api's. Also these transformations require searching for a huge combinations tree of components, and might be some specification exists that are not required so these phase is responsible to check that each specification transformation is required to build the scene, for example the user might put for the same data source a file path and file url, and one of them are necessary.\\

After builing phase connecting is required, and here's where linking phase responsibility comes. Linking main responsibility is to connect layers to each other, for example axis and geometry layers requires scale layer, so the \libname~ requires to traverse for all user's specification and search for each node connection and connect/link it for it's requirements. Another responsibility is to check for linking or connecting acceptability, not all layers accept to connect to each other, so \libname~ require to check if each user's connection is accepted or not.\\

Last thing last is assemble phase, where the user will see the result of his/her hand. assemble phase take all transformed specification that turned to layers functions/api and execute it to the selected HTML5 Canvas, the phase is similar to code generating and optimizing where it takes each layer's functions and put them in a queue to be executed in the same order, for example, we have to execute scales before axis, and run color palettes before geometries, and so on.

\subsection{\libname\space from \textbf{H}uman \textbf{C}omputer \textbf{I}nteraction perspective}
\subsubsection{Discoverability}
From (Norman, 2002)'s point of view, \textit{discoverability} is to figure out possible actions and
how to do them. Meanwhile, (Nielsen, 1994) suggests that \textit{discoverability} is minimizing the
user's memory load by making objects, actions, and options visible as possible. Instructions for the
use of the system should be visible or easily retrievable. In \libname, layers are understood by
their name (i.e., geometries layer clearly read that it is for generating geometrical objects in a
scene).

\subsubsection{Mapping}
\cite{Norman} believes that \textit{mapping} is a technical term which means the relationship
between two instances of things (data and its visual representation in our case). On the other hand,
\cite{Nielsen} says that a system would exhibit \textit{mapping} if it speaks the users' language,
with words, phrases, and concepts familiar to the user, rather than system-oriented terms. Follow
real-world conventions, making information appear in a natural and logical order. As \libname\space
is a visualization library, its foremost concern is mapping data to graphics, using same words and
phrases the other libraries apply, making understanding it or switching from any library a soft
touch.

\subsubsection{Affordance}
One way to make the interface both manageable and usable is to design interfaces that by their very
design inform users how to. \cite{Norman} defined \textit{affordance} to be a relationship between
the properties of an object and the capabilities of the agent that determine how the object could be
possibly used. As illustrated earlier, \libname\space uses keywords and functions which tell the
user how it operates.

\subsubsection{Structure}
As \cite{Constantine} proposes, a software would employ the \textit{structure} principle if its
user's interface design is organized purposefully, in meaningful and useful ways based on precise,
consistent modes that are apparent and recognizable to users, putting related things together and
separating unrelated things, differentiating different things and making similar things resemble one
another. \libname's user grammar based components which come from internal layers provide a
structure by itself that helps to manage the overall graphical scene generation.

\subsubsection{Ease/Comfort}
Ease and Comfort are two similar ideas come from the principles of universal design,
\cite{uni_design} defined \textit{ease} as using software efficiently, comfortably and with a minimum
of fatigue. While defining \textit{comfort} as presenting appropriate size and space for approach,
reach, manipulation, and use regardless of user’s body size, posture, or mobility. It is quite
apparent that these two concepts achieved in \libname\space because of the compiler as illustrated
previously.

\subsubsection{Flexibility}
Defined by \cite{Nielsen} as speeding up the interaction for the expert user such that the system
can cater to both inexperienced and experienced users. Allowing users to tailor frequent
actions. \cite{uni_design} represents it as either the design accommodated a wide range of individual
preferences and abilities or not. \libname\space obeys both definitions since it allows its users to
generate the same plot in many ways, hence making proper for both novice and expert users from
various domains.



\lstset{
  string=[s]{"}{"},
  stringstyle=\color{blue},
  comment=[s]{:\ "}{"},
  commentstyle=\color{red},
  basicstyle=\footnotesize,
  frame = single,
  framexleftmargin=0pt,
  framexrightmargin=-10pt,
}

\section{Examples}\label{sec:exampl-results-disc}
\subsection{Parallel Coordinates Example}
\begin{figure}[tbh]
  \centering
  \includegraphics[width=0.5\textwidth]{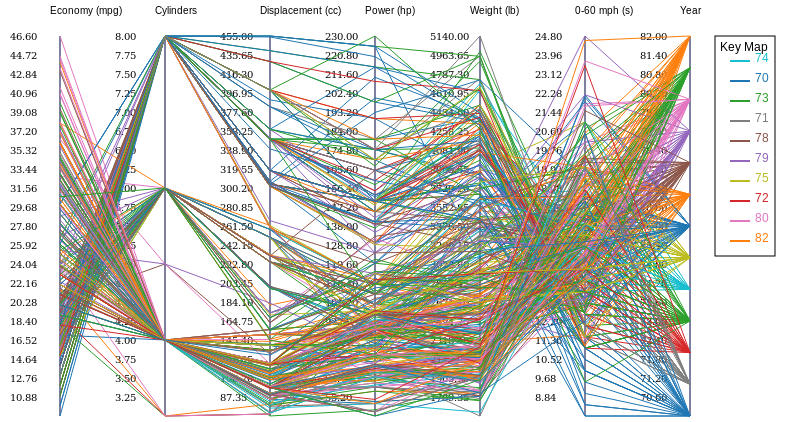}
  \caption{Caption place holder}\label{fig:parallel_cooridnates}
\end{figure}
Figure~\ref{fig:parallel_cooridnates} presents the benchmarking results, as illustrated in the previous
section, for all \libname\space visualization, the time it takes starting from page loading to
rendering scene graphs typically faster than most of the libraries; the main reason is the
\textit{HTML5 Canvas} built on top of WebGL that uses graphics card as a computational booster.

\bigskip

First and foremost, we have to load the data variable, CSV type in this situation (For JSON, specify
the type to be \textit{"json"}). We can import from multiple data sources and execute some
operation on them using \textit{transformation} and \textit{statistics} layers.
\begin{lstlisting}
"data": [{
  "name": "crimea",
  "values": "crimea-parallel.csv",
  "format": {
    "type": "csv"
  }
}]
\end{lstlisting}
After finishing data loading, the \textit{transformation} phase begins, since we do not need it in the situation, we specify it to be empty.
\begin{lstlisting}
"transform": []
\end{lstlisting}

We then move to set \textit{scales} used for the axes, in this example, we will use the \textit{linear} scale as the data does not need scaling. We can use as many scales as we want.
\begin{minipage}[t]{0.25\textwidth}
\begin{lstlisting}
"scales": {
  "name": "xscale",
  "type": "linear",
  "range": {
    "type": "range",
    "value": [0, 330]
  },
  "domain": {
    "data": "crimea",
    "field": "name"
  }
}
\end{lstlisting}
\end{minipage}%
\begin{minipage}[t]{0.25\textwidth}
\begin{lstlisting}
"encoding": {
  "x": {
    "field": "wavelength",
    "type": "quantitative",
    "scale": {
      "domain": [300,450]
    }
  },
  "y": {
    "field": "power",
    "type": "quantitative"
  }
}
\end{lstlisting}
\end{minipage}
Defining \textit{axes} follows placing scales. An axes must have some characteristics such as:
\begin{itemize}
  \item \texttt{Type:} the type in the snippet below implemented uniquely for polar parallel coordinates. There are tons of axes types.
  \item \texttt{Properties:} define aesthetics of the axes; annotation sets axes' title text, position, and color.
  \item \texttt{Transform:} sets the transformation of axes, in our case, the axes uses the \textit{power function} transformation.
\end{itemize}
And as you see in comparison code snippets, the difference between \libname~ and \textbf{Vega-Lite}, you can see that vega-lite for each variable you define it's scale or uses the default scale for the type(quantitative in example), but if you are using \libname~ you only need to define the scale and use it's name to invoke it on as many variables/axes as you want, and here you can sense the power of dynamicity.

\begin{minipage}{0.3\textwidth}
  \lstset{framexrightmargin=-2pt, basicstyle=\footnotesize}
\begin{lstlisting}{}
"axes": {
  "type": "coord_polar_parallel",
  "properties": [{
     "type": "y",
     "data": "crimea",
     "field": "economy-mpg",
     "orient": "left",
     "grid": false,
     "text": {
       "font": "10px tohma",
       "colour": "blue"
     },
     "annotation": {
       "title": "economy-mpg-",
       "position":
          "edge",
       "font": "10px Arial",
       "colour": " blue"
     },
     "transform": {
       "function": "pow",
       "properties": {
         "power": 2,
         "name": "power"
	}
  }]
}
\end{lstlisting}
\end{minipage}%
\begin{minipage}{0.20\textwidth}
  \begin{figure}[H]
    \includegraphics[width=\textwidth]{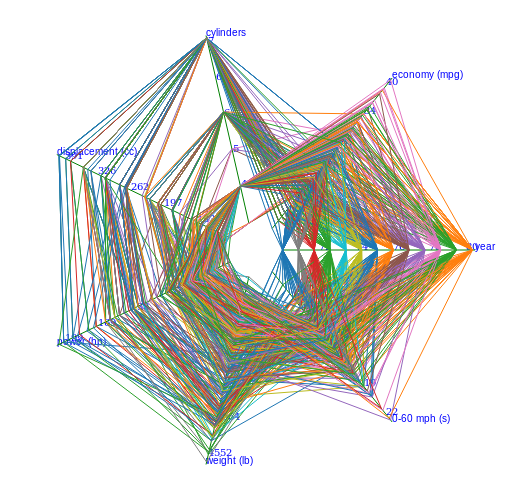}
  \end{figure}
  \begin{figure}[H]
    \includegraphics[width=\textwidth]{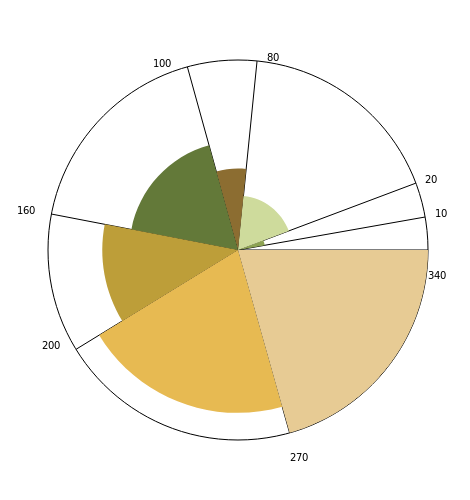}
  \end{figure}
\end{minipage}
The
The figures above reveal the power of using \libname, both are \cite{parallel_coord}'s parallel coordinates, but the difference as shown is that one of them is polar which is a compelling point in the library, that you can easily convert any visualization
scene from cartesian to polar coordinate in a single command.

%


\section{Conclusion}\label{sec:conclusion}
We have presented the JavaScript Open-source Library (\libname), the grammar of graphics JavaScript library which has the power to
generate the most captivating plots with no limitations. \libname\space is implemented using
JavaScript, which enabled it to be integrable to many other fields. We also demonstrated how it
works internally by describing its layers' specification, role, and the architectural design. We
discussed how \libname\space accompanies the Human-Computer Interaction (HCI) principles. Manipulating data is a crucial part of \libname, the way data loaded from various sources, and how transformation (e.g., filtering and grouping) and statistical (e.g., mean and std)
operations are applied, the way of generating scales for mapping data fed to the geometries layer
proffered \libname\space an edge while comparing to other libraries.  Another contribution of the
library is the way of combining low-level visualization layer with the compilation process. And we
considered a comparative study between \libname\space and great libraries such \cite{2011-d3} and
\cite{ggplot2}, and a comprehensive comparison with \cite{vega-lite}. Finally, we explained how to
use \libname\space through numerous examples such as the parallel plot in Cartesian and Polar
scales.



\if 0\MYJOURNAL%
    \bibliographystyle{elsarticle-num}
    \section*{References}
    \bibliography{citations}
\else \if 1\MYJOURNAL%
    \small
    \bibliographystyle{IEEEtran}
    \bibliography{citations}

\begin{thebibliography}{10}
\providecommand{\url}[1]{#1}
\csname url@samestyle\endcsname
\providecommand{\newblock}{\relax}
\providecommand{\bibinfo}[2]{#2}
\providecommand{\BIBentrySTDinterwordspacing}{\spaceskip=0pt\relax}
\providecommand{\BIBentryALTinterwordstretchfactor}{4}
\providecommand{\BIBentryALTinterwordspacing}{\spaceskip=\fontdimen2\font plus
\BIBentryALTinterwordstretchfactor\fontdimen3\font minus
  \fontdimen4\font\relax}
\providecommand{\BIBforeignlanguage}[2]{{%
\expandafter\ifx\csname l@#1\endcsname\relax
\typeout{** WARNING: IEEEtran.bst: No hyphenation pattern has been}%
\typeout{** loaded for the language `#1'. Using the pattern for}%
\typeout{** the default language instead.}%
\else
\language=\csname l@#1\endcsname
\fi
#2}}
\providecommand{\BIBdecl}{\relax}
\BIBdecl

\bibitem{JSOL2016}
\BIBentryALTinterwordspacing
W.~A. Yousef, H.~E. Mohammed, A.~A. Naguib, Y.~M. Khalifa, A.~M. Mamdoh, E.~A.
  Awad, N.~A. S. S.~T. AbdElrheem, and S.~G. Gaafar, ``Jsol: Javascript-based
  open-source library,'' 2016. [Online]. Available:
  \url{https://github.com/hci-lab/JSOL}
\BIBentrySTDinterwordspacing

\bibitem{Yousef2019DVP-arxiv}
W.~A. Yousef, A.~A. Abouelkahire, O.~S. Marzouk, S.~K. Mohamed, and M.~N.
  Alaggan, ``{DVP: Data Visualization Platform},'' \emph{arXiv preprint
  arXiv:1906.11738}, 2019.

\bibitem{2011-d3}
\BIBentryALTinterwordspacing
M.~Bostock, V.~Ogievetsky, and J.~Heer, ``D3: Data-driven documents,''
  \emph{IEEE Trans. Visualization \& Comp. Graphics (Proc. InfoVis)}, 2011.
  [Online]. Available: \url{http://vis.stanford.edu/papers/d3}
\BIBentrySTDinterwordspacing

\bibitem{Bostock2009Protovis}
M.~Bostock and J.~Heer, ``Protovis: A graphical toolkit for visualization,''
  \emph{IEEE transactions on visualization and computer graphics}, vol.~15,
  no.~6, 2009.

\bibitem{ggplot2}
\BIBentryALTinterwordspacing
H.~Wickham, \emph{ggplot2: Elegant Graphics for Data Analysis}.\hskip 1em plus
  0.5em minus 0.4em\relax Springer-Verlag New York, 2009. [Online]. Available:
  \url{http://ggplot2.org}
\BIBentrySTDinterwordspacing

\bibitem{vega-lite}
A.~Satyanarayan, D.~Moritz, K.~Wongsuphasawat, and J.~Heer, ``Vega-lite: a
  grammar of interactive graphics,'' \emph{IEEE Transactions on Visualization
  and Computer Graphics}, vol.~23, no.~1, pp. 341--350, 2017.

\bibitem{tableau}
C.~Lutz, F.~Wolter, and M.~Zakharyaschev, ``A tableau algorithm for reasoning
  about concepts and similarity,'' in \emph{Automated Reasoning with Analytic
  Tableaux and Related Methods}, M.~Cialdea~Mayer and F.~Pirri, Eds.\hskip 1em
  plus 0.5em minus 0.4em\relax Berlin, Heidelberg: Springer Berlin Heidelberg,
  2003, pp. 134--149.

\bibitem{Wilkinson}
L.~Wilkinson, \emph{The Grammar of Graphics (Statistics and Computing)}.\hskip
  1em plus 0.5em minus 0.4em\relax Secaucus, NJ, USA: Springer-Verlag New York,
  Inc., 2005.

\bibitem{Polaris}
\BIBentryALTinterwordspacing
C.~Stolte, D.~Tang, and P.~Hanrahan, ``Polaris: a system for query, analysis,
  and visualization of multidimensional databases,'' \emph{Commun. ACM},
  vol.~51, no.~11, pp. 75--84, Nov. 2008. [Online]. Available:
  \url{https://doi.org/10.1145/1400214.1400234}
\BIBentrySTDinterwordspacing

\bibitem{bellman}
R.~E. Bellman, \emph{Adaptive Control Processes}.\hskip 1em plus 0.5em minus
  0.4em\relax Princeton, NJ: Princeton University Press, 1961.

\bibitem{Norman}
D.~A. Norman, \emph{The Design of Everyday Things}.\hskip 1em plus 0.5em minus
  0.4em\relax New York, NY, USA: Basic Books, Inc., 2002.

\bibitem{Nielsen}
\BIBentryALTinterwordspacing
J.~Nielsen, ``Usability inspection methods,'' in \emph{Conference Companion on
  Human Factors in Computing Systems}, ser. CHI '94.\hskip 1em plus 0.5em minus
  0.4em\relax New York, NY, USA: ACM, 1994, pp. 413--414. [Online]. Available:
  \url{https://doi.org/10.1145/259963.260531}
\BIBentrySTDinterwordspacing

\bibitem{Constantine}
L.~L. Constantine and L.~A.~D. Lockwood, \emph{Software for Use: A Practical
  Guide to the Models and Methods of Usage-centered Design}.\hskip 1em plus
  0.5em minus 0.4em\relax New York, NY, USA: ACM Press/Addison-Wesley
  Publishing Co., 1999.

\bibitem{uni_design}
R.~L. M.~F. Molly Follette~Story, James L.~Mueller, \emph{The Universal Design
  File: Designing for People of All Ages and Abilities. Revised Edition.},
  1998.

\bibitem{parallel_coord}
\BIBentryALTinterwordspacing
A.~Inselberg and B.~Dimsdale, ``Parallel coordinates: A tool for visualizing
  multi-dimensional geometry,'' in \emph{Proceedings of the 1st Conference on
  Visualization '90}, ser. VIS '90.\hskip 1em plus 0.5em minus 0.4em\relax Los
  Alamitos, CA, USA: IEEE Computer Society Press, 1990, pp. 361--378. [Online].
  Available: \url{http://dl.acm.org/citation.cfm?id=949531.949588}
\BIBentrySTDinterwordspacing

\end{thebibliography}
    \normalsize
    \fi
\fi

\end{document}